\documentclass[10pt,letterpaper]{article}
\usepackage{opex3}
\usepackage{amssymb}
\usepackage{amsmath}
\usepackage{graphicx,float}
 \usepackage{subfigure}
 \usepackage{dcolumn}
\usepackage{bm}
\usepackage{mathrsfs}
\usepackage{txfonts}
\usepackage{cite}

\usepackage{epsfig}
\usepackage{color}

\begin{document}

\title{Deterministic generation of an on-demand Fock state}


\author{Keyu Xia,$^{1,*}$ Gavin K. Brennen,$^{1}$ Demosthenes Ellinas,$^{2}$ and Jason Twamley$^{1}$}

\address{
$^1$Centre for Engineered Quantum Systems, Department of Physics and Astronomy, \\Macquarie University, NSW 2109, Australia\\
$^2$Technical University of Crete, 
 Department of Sciences M$\Phi$Q Research Unit,\\
 GR-731 00 Chania, Crete Greece
}

\email{$^*$keyu.xia@mq.edu.au} 




\begin{abstract} We theoretically study the deterministic generation of photon Fock states on-demand using a protocol based on a Jaynes Cummings quantum random walk which includes damping. We then show how each of the steps of this protocol can be implemented in a low temperature solid-state quantum system with a Nitrogen-Vacancy centre in a nano-diamond coupled to a nearby high-Q optical cavity. By controlling the coupling duration between the NV and the cavity via the application of a time dependent Stark shift, and by increasing the decay rate of the NV via stimulated emission depletion (STED) a Fock state with high photon number can be generated on-demand. Our setup can be integrated on a chip and can be accurately controlled.\end{abstract}


\bibliographystyle{osajnl}


\section{Introduction}
The generation of on-demand photonic Fock states is at the heart of many photonic quantum technologies. Single-photon sources have been realized in a variety of quantum systems such as nitrogen-vacancy (NV) centres in diamond \cite{NaturePhoton5p738}, or using quantum dots \cite{Rivoire2011}. 
However the creation of photonic Fock states with a high photon number is an open challenge to date. In this paper we propose a novel Jaynes Cummings quantum random walk (QRW) protocol that drives the cavity to accumulate a photonic Fock state deterministically. We describe in detail how to implement the theoretical protocol using a Nitrogen-Vacancy defect in a nano-diamond evanescently coupled to circulating light modes in a high-Q toroidal resonator at moderately low temperatures ($<10$ $\;{\rm K}$). We show that even in the case where one has error in the timing of the control pulses there is very high probability to generate photonic Fock states up to $n=6$.
 
Synthesising high-number Fock states has received much attention in the literature. A high-number Fock state has been conditionally produced with a probability $P_n$ via the state collapse from a coherent or thermal state \cite{Nature448p889,PRL87p093601,PRL65p976}. The probability $P_n$ of success is equal to the initial overlap probability of the target Fock state with the initial state of light ($P(n)={\rm Tr}[ |n\rangle\langle n| \rho_{init}]$), and this probability can be quite low: for $|n=3\rangle$,  $P_3\approx 0.22$ if starting from a pure coherent state $|\alpha=\sqrt{3}\rangle$.
An arbitrary quantum state of a cavity field can also be engineered if the  cavity-qubit coupling can be very accurately controlled and recent experiments using low-temperature superconducting circuit-QED have synthesised microwave cavity states up to nine photons \cite{PRL76p1055,EPL67p941,Nature454p310}. 
When an excited two level system interacts with a cavity mode via the Jaynes Cummings (JC) interaction of strength $g$, the emission probability $P_{emit}(g,\tau,n)$ of the excited atom depends on the duration of coupling $\tau$ and the choice of Fock state $n$. For certain values, terming trapping values, of $(g, \tau,n)$,  $P_{emit}$ vanishes and a Fock state can be trapped in the cavity \cite{JOSAB3p906,PRL86p3534,NJP6p97}. In this way by sending a train of Rydberg atoms through a superconducting high-Q microwave cavity, Walther {\it et al.} trapped a microwave Fock state  \cite{PRL86p3534,NJP6p97}.  Indeed all the experimental demonstrations in high-Fock number state generation  have been in the microwave regime \cite{Nature454p310,PRL86p3534,NJP6p97}. At optical frequencies, Brown {\it et al.} propose a system of $N$ three-level atoms in a high-finesse cavity \cite{PRA67p043818}, for Fock state generation but this requires the preparation of complicated nonclassical states of the atoms. Until now only single photon sources at optical frequencies have been realized in solid-state quantum systems.

{\it Jaynes Cummings Damped Quantum Random Walk:-} A coined quantum random walk involves a coin, which we take as a qubit  with Hilbert space
$\mathcal{H}_{c}=span\{\left\vert e\right\rangle,\left\vert g\right\rangle\}$, together with a walk on the discretised non-negative real line 
$\mathcal{H}_{w}=span\{\left\vert
n\right\rangle ;n=0,1,\cdots\}$. The normal coined quantum random walk on the full real line ($-\infty\le n \le \infty$), is an iteration of a basic step involving a conditional displacement of the walker on the line depending on the internal state of the walker $\hat{U}_d\equiv  |e\rangle\langle e|\otimes |n+1\rangle\langle n|+ |g\rangle\langle g|\otimes |n-1\rangle\langle n|$, followed by a ``scrambling'' of the internal state of the walker by the action of a Hadamard operation on the internal states. This coined version of the QRW where the walker moves on the discretized real line $\mathbb{Z}$ has been studied intensively over the past decade. In the following we will examine the case when the space upon which the walker walks is the Fock ladder, $n\in \mathbb{Z}^*$, i.e. the non-negative integers. It is no longer possible to have a unitary operator that implements a conditional displacement with constant displacement independent of the position of the walker. To achieve a unitary operation for the conditional displacement the walker can execute a step up/down the half-line with ``step sizes'' that depend on $n$. Our QRW step will consist of a period of Jaynes Cummings evolution between the internal states of the coin and conditional displacements up/down the Fock ladder, followed by a manipulation of the internal states of the walker. Rather than a complete scrambling of the internal states we will just consider a flip where $|g\rangle \leftrightarrow |e\rangle$, are swapped. We have found that such a unitary QRW on the half line using the JC walk step exhibits complex temporal dynamics but simplifies greatly when we allow for periodic damping of the internal state of the coin.

We now consider analytically the above QRW on the half-line and derive a formulae for the resulting map on the reduced Fock space of the walker.  We will see that, if starting at $|n=0\rangle$, the walker will, on average, step to greater values of  $n$ and will hit a ceiling value of $n$ that depends on the chosen value for the JC interaction strength/time. Let the JC\ Hamiltonian in the RWA be given by $H_{JC}=g(\left\vert e\right\rangle
\left\langle g\right\vert \otimes \hat{a}+\left\vert g\right\rangle \left\langle
e\right\vert \otimes \hat{a}^{\dagger }),$ then the resulting unitary evolution operator 
$\hat{U}_{JC}(\tau)=e^{-iH_{JC}\tau},$ can be expressed as
\begin{equation}
\hat{U}_{JC}(\tau)=\left( 
\begin{array}{cc}
\cos g\tau\sqrt{N+1} & -i\frac{\sin (g\tau\sqrt{N+1})}{\sqrt{N+1}}a \\ 
-ia^{\dagger} \frac{\sin (g\tau\sqrt{N+1})}{\sqrt{N+1}} & \cos g\tau\sqrt{N}%
\end{array}%
\right)\;\; ,
\end{equation}
where $N=a^\dag a$ is the photon number operator.
Considering the initial product state for the density matrix of the coin and the walker to be $\rho _{C}\otimes \rho _{W}$, then 
following evolution by the JC Hamiltonian we obtain $\rho _{C}\otimes \rho _{W}\rightarrow
\hat{U}_{JC}\,(\rho _{C}\otimes \rho _{W})\,\hat{U}_{JC}^{\dagger }$. Subsequently we allow
spontaneous emission (amplitude damping channel) to operate on the atomic
system i.e. \ 
\begin{equation}
\rho _{C}\otimes \rho _{W}\rightarrow \hat{U}_{JC}\,(\rho _{C}\otimes \rho
_{W})\,\hat{U}_{JC}^{\dagger }\rightarrow  \hat{\mathcal{E}}_{SE}\otimes \hat{id}_{W}\,\left[\hat{U}_{JC}\,(\rho
_{C}\otimes \rho _{W})\,\hat{U}_{JC}^{\dagger }\right] ,
\end{equation}%
where $\hat{id}_{W}$ stands for the identity map in walker's space.
Here $\hat{\mathcal{E}}_{SE}=Ad\,\hat{S}_{0}+Ad\,\hat{S}_{1}$ is the spontaneous emission channel
with non-unitary Kraus generators%
\begin{equation}
\hat{S}_{0}=\left\vert g\right\rangle \left\langle g\right\vert +\left\vert
e\right\rangle \left\langle e\right\vert \sqrt{\eta },\;\;%
\hat{S}_{1}=\left\vert g\right\rangle \left\langle e\right\vert \sqrt{1-\eta },
\end{equation}%
where $\eta (t)=e^{-t/T}$ a positive parameter quantifying the degree by which the atomic system is reset by the channel, with 
$t$ the nominal time over which the channel operates and $T$ a constant characterising how rapid the reset process it.
We have also used the notation of the adjoint action $Ad\,(\hat{A})$ of an operator $\hat{A}$ on some
other operator $\hat{X}$ as follows: \ $\hat{X}\rightarrow Ad(\hat{A})\hat{X}=\hat{A}\hat{X}\hat{A}^{\dagger },$
noticing the property $Ad(\hat{A}\hat{B})\hat{X}=Ad(\hat{A})Ad(\hat{B})\hat{X}.$
For a general pure input state of the coin: 
\begin{equation}
\rho _{C}=(\alpha \left\vert g\right\rangle +\beta \left\vert e\right\rangle
)(\alpha ^{\ast }\left\langle g\right\vert +\beta ^{\ast }\left\langle
e\right\vert )=\left( 
\begin{array}{cc}
\left\vert \beta \right\vert ^{2} & \alpha^{\ast} \beta  \\ 
\alpha \beta^{\ast}  & \left\vert \alpha \right\vert ^{2}%
\end{array}%
\right) ,
\end{equation}
the channel $\hat{\mathcal{E}}_{SE}$ outputs 
\begin{equation}
\hat{\mathcal{E}}_{SE}[\rho _{C}](t)=\left( 
\begin{array}{cc}
\eta \left\vert \beta \right\vert ^{2} &~~~ \alpha^{\ast}\beta \sqrt{%
\eta } \\ 
\alpha \beta^{\ast} \sqrt{\eta } &~~~ 1-\eta \left\vert \beta \right\vert ^{2}%
\end{array}%
\right) .
\end{equation}
In view of the limit $\lim_{t\rightarrow \infty }\eta (t)=0,$ and
normalization relation $\left\vert \alpha \right\vert ^{2}+\left\vert \beta
\right\vert ^{2}=1,$ the last expression leads to the reset state $%
\lim_{t\rightarrow \infty }\hat{\mathcal{E}}_{SE}[\rho _{at}](t)=\left\vert
g\right\rangle \left\langle g\right\vert =\left( 
\begin{array}{cc}
{\scriptsize 0} & {\scriptsize 0} \\ 
{\scriptsize 0} & {\scriptsize 1}%
\end{array}%
\right) $. 

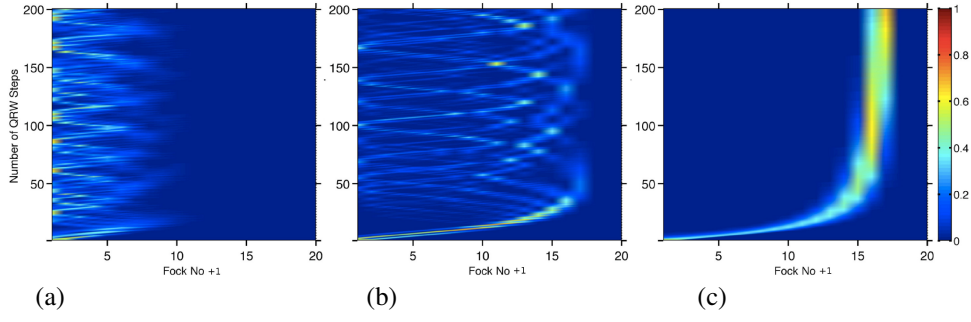
\begin{figure}
\begin{center}
\setlength{\unitlength}{1cm}
\begin{picture}(8.5,4)
\put(-2.3,0){\includegraphics[width=13cm]{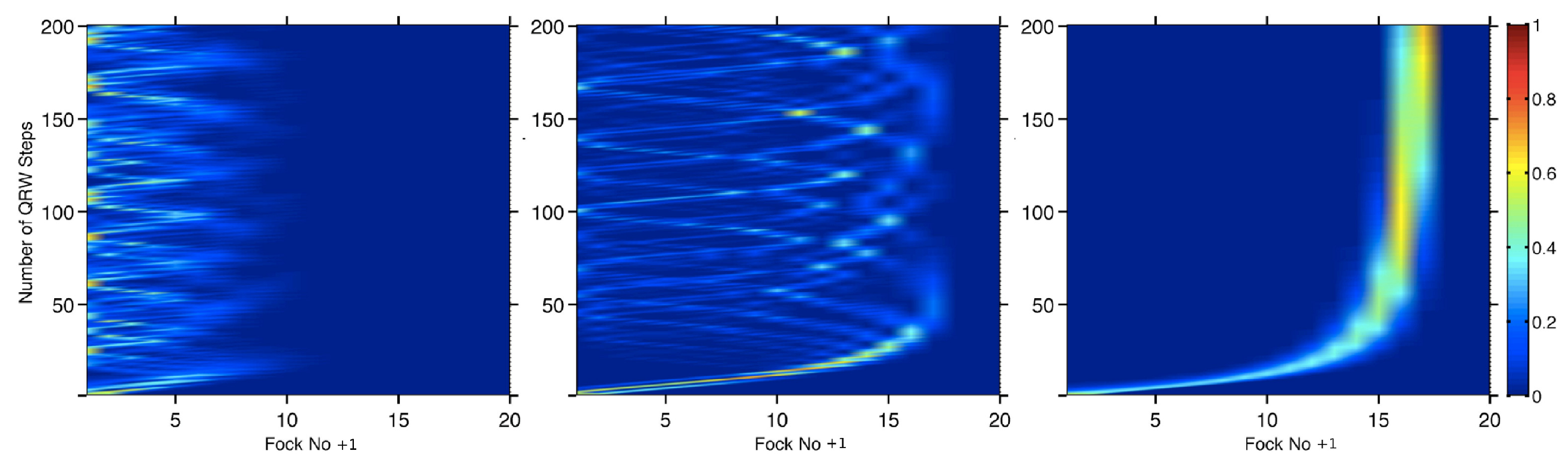}}
\put(-1.8,-.3){(a)}
\put(2.6,-.3){(b)}
\put(7,-.3){(c)}
\end{picture}
\end{center}
\caption{Jaynes Cummings quantum random walk:  Plots showing how the walker evolves starting in the vacuum, i.e. ${\rm Tr}\{ |n\rangle\langle n|\, \hat{\cal E}^m[ |e\rangle\langle e|\otimes |0\rangle\langle 0|]\}$, as a function of the number of steps $m$ and Fock number $n$. (a) for a completely unitary evolution with $\hat{\cal E}=Ad\hat{H}\circ Ad\hat{U}_{JC}$, where one executes a Hadamard on the coin space (b) completely unitary evolution with $\hat{\cal E}=Ad\hat{X}\circ Ad\hat{U}_{JC}$, where one executes a $\pi$ flip instead of the Hadamard and (c) including spontaneous damping channel $\hat{\cal E}=Ad\hat{X}\circ Ad\hat{\cal E}_{SE}\circ Ad\hat{U}_{JC}$, and the Jaynes Cumming coupling strength and duration chosen so that $|n=16\rangle$ is a trapping state. We see that the latter evolution clearly leads to accumulation of the walker at the target trapping state.}\label{fig:QRW}
\end{figure}

Next we form the composite map beginning with the JC unitary, spontaneous emission channel  and then finally a flipping of the atomic state $\hat{X}\equiv \exp(-i\pi/2 \sigma_x)$ and denote the entire process by $\hat{\mathcal{E}}$:
\begin{equation}
\hat{\mathcal{E}}\equiv \;\;Ad\hat{X}\circ (\hat{\mathcal{E}}_{SE}\otimes \hat{id}_{W})\circ
Ad\hat{U}_{JC},
\label{map}
\end{equation}%
 where $\hat{\mathcal{E}}$ \ acts in total coin-walker (atom-mode) density matrix.
Choosing the action $\hat{\mathcal{E}}(\left\vert e\right\rangle \left\langle
e\right\vert \otimes \rho _{W}\mathcal{)},$ gives \cite{ellinas2005}, 
\begin{eqnarray}
&&\hat{\mathcal{E}}(\left\vert e\right\rangle \left\langle e\right\vert \otimes
\rho _{W}\mathcal{)}  \nonumber \\
&=&\left( 
\begin{array}{cc}
\hat{\mathcal{E}}_{W}(\rho _{W})-\cos (g\tau\sqrt{N+1})\rho _{W}\cos (g\tau\sqrt{N+1}%
)\eta  & ia^{\dagger }\frac{\sin (g\tau\sqrt{N+1})}{\sqrt{N+1}}\rho _{W}\cos (g\tau%
\sqrt{N+1})\sqrt{\eta } \\ 
-i\cos (g\tau\sqrt{N+1})\rho _{W}\frac{\sin (g\tau\sqrt{N+1})}{\sqrt{N+1}}a\sqrt{%
\eta } & \cos (g\tau\sqrt{N+1})\rho _{W}\cos (g\tau\sqrt{N+1})\eta 
\end{array}%
\right) ,  \nonumber \\
&&
\end{eqnarray}%
where the positive map%
\begin{equation}
\hat{\mathcal{E}}_{W}(\rho _{W})=\cos (g\tau\sqrt{N+1})\rho _{W}\cos (g\tau\sqrt{N+1}%
)+a^{\dagger }\frac{\sin (g\tau\sqrt{N+1})}{\sqrt{N+1}}\rho _{W}\frac{\sin (g\tau%
\sqrt{N+1})}{\sqrt{N+1}}a,  \label{cpmapw}
\end{equation}%
appearing above can be shown to be trace preserving i.e. 
\begin{equation}
{\rm Tr}\hat{\mathcal{E}}_{W}(\rho _{W})={\rm Tr}\{[\cos (g\tau \sqrt{N+1})^{2}+\sin (g\tau \sqrt{N+1}%
)^{2}]\,\rho _{W}\}={\rm Tr}\rho _{W}=1.
\end{equation}
The model
described leads to a sequence of walker's density matrices \cite{bracken2004}, and after $m>1$ steps the reduced state of the walker is
\begin{equation}
\rho _{W}^{(m)}={\rm Tr}_{c}[\hat{\mathcal{E}}^{m}\mathcal{(}\left\vert e\right\rangle \left\langle
e\right\vert \otimes \rho _{W}\mathcal{)}]= \hat{\mathcal{E}}_{W}^{m}(\rho
_{W})+\mathcal{O}(\eta ^{\frac{3}{2}}).
\end{equation}
Consider the case of number state input for the walker $\rho _{W}=\left\vert n\right\rangle \left\langle
n\right\vert $.  Then
\begin{equation}
\hat{\mathcal{E}}_{W}(\left\vert n\right\rangle \left\langle n\right\vert )=
\cos (g\tau\sqrt{n+1})^{2}\left\vert n\right\rangle \left\langle n\right\vert
+\sin (g\tau\sqrt{n+1})^{2}\left\vert n+1\right\rangle \left\langle
n+1\right\vert,
\end{equation}
and repeated action of $\hat{\mathcal{E}}$ in the $\eta \rightarrow 0$ limit
leads to a progressive increase in Fock number $n$.
From the form of $\hat{\mathcal{E}}_{W}$ in this limit we now observe that we can halt this upwards motion to accumulate at the trapping value $n=n_T$, if we choose the Jaynes Cummings coupling strength and duration  $\tau$ to satisfy  the trapping condition: $g\tau\sqrt{n_T+1}=k\,\pi$, where $k\in\mathbb{Z}$. {\em This is the main result of this section: by executing a sequence of operations: Jaynes Cummings for a period, followed by spontaneous decay and then a complete flip of the coin space, one can, with unit probability, arrange for the walker to reach a steady state at the position $n=n_T$}.
The behaviour of the quantum random walk on the half line with and without damping can be observed in Fig. \ref{fig:QRW}.  The dramatically different behaviour from the position independent step operator used in the conventional quantum walk is apparent.  Next we show how to implement this map for an optical cavity-QED setup consisting of a Nitrogen-Vacancy centre in a nano diamond coupled to a high-Q optical cavity to produce multi-photon optical Fock states. {  We perform more detailed numerical simulations taking into account cavity and atomic decay to determine a figure of merit to synthesis Fock states of light.}
\\

Now we theoretically discuss the maximum Fock number that can be reached with high fidelity. We consider a case where the noise in the timing of the JC interactions is vanishingly small. However the decay of cavity and the effect of population in the ground state of the qubit must be taken into account. We denote the effective cavity loss of the photons with the rate $\gamma_c$. The higher the Fock state is, the larger this effective decay rate. For the target Fock state $n_T$, the effective decay rate increases to $n_T\gamma_c$. Another factor limiting whether one can achieve the target state relates to the downward transfer of population with probability $P_D$ from the target state $|n_T\rangle$ to the lower Fock state $|n_T -1\rangle$ due to the net population $P_g$ of the ground state. In the stationary state, the pumping  probability $P_U$ from the state $|n_T-1\rangle$ must balance the loss from the target Fock state $|n_T\rangle$. A formula describing this balance takes the form
\begin{equation}
 P_{n_T} \left( 1- e^{-n_T\gamma_c t}\right) + P_gP_D=P_{n_T-1}P_U \,,
\end{equation}
where $P_{n_T} (P_{n_T-1})$ is the population in Fock state $|n_T \rangle (|n_T -1\rangle)$. Because $\eta$ can not be practically zero after waiting for a time $t$, the net population in the excited state of qubit is $\eta=e^{-t\gamma_q}$ with the effective decay rate of qubit $\gamma_q$ which can be modified using STED beam in our setup. After the state flipping, this population is transferred to the ground state. If the time $t$ is measured as $t=M\gamma_q^{-1}$, $P_g=e^{-M}$. We are interested in the case of high fidelity $F$ of achieving the target state. We observe that the population is a good approximation of fidelity, $P_{n_T} \sim F$. The population of  the lower Fock state is $P_{n_T-1} = \alpha (1-F)$ with constant $0\leq \alpha \leq 1$. For our Hermitian system, we have $P_U=P_D=\sin^2\left(\pi \frac{\sqrt{n_T}}{\sqrt{n_T+1}}\right)$. There we have
\begin{equation}
 F(1-e^{-N_T \gamma_c M \gamma_q^{-1}}) + e^{-M} \sin^2 \left( \pi \frac{\sqrt{n_T}}{\sqrt{n_T+1}} \right) =\alpha (1-F)\sin^2\left( \pi \frac{\sqrt{n_T}}{\sqrt{n_T+1}}\right)\,.
\end{equation}
This formula shows the relation between the decay rates, target state Fock number and the achievable fidelity. Assuming that $N_T \gamma_c M \gamma_q^{-1} \ll 1$ and $\frac{\sqrt{n_T}}{\sqrt{n_T+1}} \approx 1$, the fidelity as a function of the decay rates and the photon number $n_T$ takes the form
\begin{equation}\label{eq:FN}
 F=\frac{\pi^2 \left(\alpha - e^{-M}\right)}{\pi^2 \alpha + 4M n_T^3 \gamma_c/\gamma_q}\,.
\end{equation}

{\it Implementation:-} To implement the above protocol we propose to use a single nitrogen-vacancy (NV$^-$) center in a nanodiamond coupled to a high-finesse toroidal optical cavity at low temperature, while the latter is also connected to an optical interferometer and where the NV's optical transition is initialised via optical pumping, brought in/out of resonance with the cavity via Stark shift tuning resulting from an electric field, and undergoes periodic optical $\pi$ flips via resonant optical laser pulses.
In more detail: when the cavity interacts on-resonance with the zero-phonon line (ZPL) of the single NV center, the de-excitation probability of the NV (treated as a two level system (TLS)),  [and consequently excitation probability of the cavity], is given by $P_{emit}(g,\tau,n)=\sin^2 (\sqrt{n+1}g\tau)$, where $n$ is the number of photons in the cavity, $g$ is the JC coupling strength, and $\tau $ is the interaction time. Choosing $\tau=\tau_T$ such that the $n_T$ photon is trapped in the cavity  we have $P_{emit}(g,\tau_T,n_T)=0$. Using a fixed $\tau_T$ as the time step in the damped JCQRW above and starting the cavity in the vacuum state leads to the cavity field undergoing a deterministic ratchet-like increase in Fock number until it accumulates at $n=n_T$.  The  trapping-state condition means that the field in the cavity reaches an upper bound and is prevented from being excited to a higher photonic number state. Thus via a precise control of the Jaynes Cummings coupling $\tau_T$, an on-demand Fock state can be deterministically trapped in the cavity starting from the vacuum state. To do this we start the following process (Eq.~(\ref{map})) from the excited state of the NV center, which is resonantly prepared by a $\pi$ laser pulse: (i) we first switch on the JC coupling by tuning off the electric field.  During this stage, the NV center emits a photon with the probability $P_g$ into the cavity. (ii) After a time $\tau_T$, the JC coupling  is turned off by bringing the NV's optical transition out of resonance with the cavity via electrical Stark control \cite{Starkshift} and the NV center is allowed to completely decay to its ground state (GS). (iii) Then we resonantly pumping the NV center to its excited state again via a $\pi$ pulse. Repeating these operations, the field in the cavity can be trapped in a selected target Fock state.

\begin{figure}
\begin{center}
\setlength{\unitlength}{1cm}
\begin{picture}(8.5,7)
\put(-1,-0.6){\includegraphics[width=9.5cm]{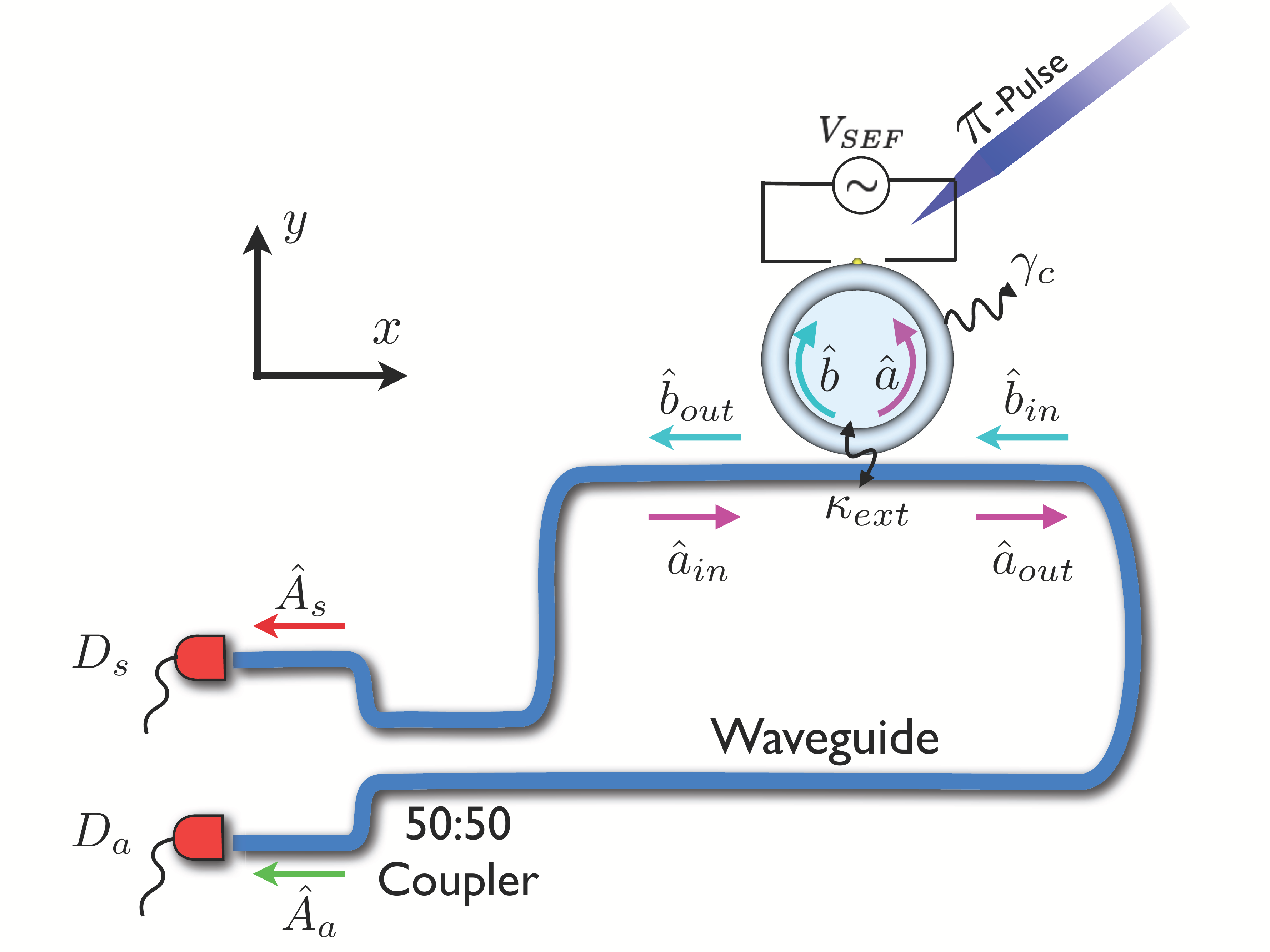}}
\end{picture}
\end{center}
\caption{Solid state setup for deterministic generation of an on-demand Fock state of photons. A two level system (nitrogen vacancy in a nanodiamond - here shown as yellow at top of the toroidal resonator), interacts with counter propagating optical modes $\hat{a}$ and $\hat{b}$ in a high-Q toroidal resonator with intrinsic decay rate $\gamma_c$ and which is coupled to a nearby waveguide interferometer at a coupling rate $\kappa_{ext}$. Shown are the input and output modes $\hat{a}/\hat{b}_{in/out}$ and the resulting anti/symmetric modes $\hat{A}_{a/s}$ modes from the interferometer with each associated photon detector $D_{a/s}$. Also shown is the incident (red arrow), laser pulse resonant on the NV zero phonon line required to implement an optical $\pi-$pulse and the Stark shift electrode used to bring the NV's optical transition in/out of resonance with the cavity. Not shown is the initialising green laser and stimulated depletion laser.}\label{fig:setup}
\end{figure}

{\it The System:-} 
Our setup for creation of photonic Fock state is shown in Fig.~\ref{fig:setup}. An optical toroidal cavity with high quality factor $Q$ and resonance frequency $\omega_c$ couples to the optical ZPL  transition in an NV$^-$ center in a type IIa nanodiamond with $C_{3v}$ symmetry, which is positioned on or nearby the toroid so that it has a large evanescent overlap with the whispering gallery optical modes of the toroidal resonator. The nanodiamond is oriented so that the $[111]$ axis of NV center is fabricated to be along the $z$ direction (see Fig.~\ref{fig:setup}). This setup can be realized using current technology \cite{APL96p241113,NanoLett9p1447,NanoLett6p2075}.   The toroidal cavity supports two degenerate modes, clockwise (CW) mode $\hat{b}$ and counterclockwise (CCW) mode $\hat{a}$, which propagate around the cavity along two opposite directions and form a standing wave if both modes were excited \cite{Science319p1062}. These two modes can be viewed in a basis of anti/symmetric modes $\hat{A}_{a,s}=(\hat{a} \mp \hat{b})/\sqrt{2}$, which are mutually orthogonal in space. They split in frequency due to the scattering $h$ from the NV center and rough surface. Depending on it's position the coupling of NV$^-$ centre to one of these two normal cavity modes $\hat{A}_{s}$ or $\hat{A}_{a}$ may occur predominantly or even exclusively with respect to the other mode  \cite{Science319p1062}.  In our setup, we take the  NV$^-$'s position  to be at the antinode of mode $\hat{A}_s$ (node of the antisymmetric mode $\hat{A}_a$). Thus the NV center couples dominantly to the mode $\hat{A}_s$. We neglect the small coupling to mode $\hat{A}_{a}$. The mode $A_s$ is set to be red detuned to the ZPL transition of the NV center in the absence of any applied Stark shift \cite{Starkshift}. The latter we propose can be used to control the coupling between the NV center and the cavity. 

An optical waveguide precisely positioned close to the cavity couples light in/out to/from the cavity with external coupling rate $\kappa_{ext}$ via the input-output relations \cite{PRA30p1386,PRA31p3761}
 \begin{align}\label{eq:InputOutput}
  \hat{a}_{out} & =-\hat{a}_{in}+ \sqrt{2\kappa_{ext}}\, \hat{a} \,,\\
  \hat{b}_{out} & =-\hat{b}_{in}+ \sqrt{2\kappa_{ext}}\, \hat{b} \,,
 \end{align}
where the input and output fields of the waveguide are denoted by $\{\hat{a}_{in},\hat{b}_{in},\hat{a}_{out},\hat{b}_{out}\}$, respectively. $[\hat{a}_{in}(t),\hat{a}_{in}^\dag(t)]=[\hat{a}_{out}(t),\hat{a}_{out}^\dag(t)]=\delta(t-t')$ and similarly $[\hat{b}_{in}(t),\hat{b}_{in}^\dag(t)]=[\hat{b}_{out}(t),\hat{b}_{out}^\dag(t)]=\delta(t-t')$. The output fields $\hat{a}_{out}$ and $\hat{b}_{out}$ are mixed by a $50:50$ directional coupler \cite{PRL105p200503}. We take the inputs to the cavity to be vacuum states, i.e. $\langle \hat{a}_{in}\rangle = \langle \hat{b}_{in}\rangle =0$, and thus the outputs $\hat{a}_{out}$ and $\hat{b}_{out}$ are proportional to $\hat{a}$ and $\hat{b}$, respectively. Thus the outputs of the directional coupler yields modes $\hat{A}_{s}$ and $\hat{A}_{a}$ \cite{splitter}, leading to the detectors. Here we aim to create Fock state of the symmetric mode $\hat{A}_s$. Assuming the intrinsic loss of the cavity is denoted by the decay rate $\gamma_c$, if we take into account the scattering $h$ between two modes $\hat{a}$ and $\hat{b}$, the critical coupling condition is given by $\kappa_{ext}=\sqrt{h^2+\gamma_c^2}$ \cite{Science319p1062}.

To switch the interaction between the NV center and the cavity, an electric field  perpendicular to the axis of NV center is applied to induce a Stark shift. This static electric field (SEF) can be created by two electrodes positioned $10$ ${\mu}{\rm m}$~above the setup \cite{PRL107p266403}. This distance is much larger than the wavelength of the field in the cavity and the extent of the evanescent field and thus results in negligible scattering loss of the cavity modes. During the excitation of cavity, this SEF is applied to shift the NV center in/out of resonance with the cavity, thus executing the JC step in the map Eq. (\ref{map}).

A critical step in the process Eq. (\ref{map}) is the rapid decay of the two level system $\hat{{\cal E}}_{SE}(\hat{\rho})$, the  spontaneous emission decay of the two level system (coin). This decay must be executed with a rate much higher than the cavity decay rate. The natural excited state lifetime of the NV ZPL is $\sim 11$ns and this is too long to permit many repetitions of our process Eq. (\ref{map}) even with high-Q cavities. To shorten this, after the JC coupling is switched off by applying the SEF we use
a stimulated emission depletion (STED) laser beam ($\lambda_{STED}=775$~${\rm nm}$), to dynamically create a fast decay channel from the excited state to the ground state of the NV center \cite{NaturePhoton3p144}, and this can increase the effective decay rate of the NV by almost four orders of magnitude. After almost all of the population has decayed to the ground state $|g\rangle$, another laser beam on-resonant with the ZPL repumps the NV center from $|g\rangle$ to $|e\rangle$ \cite{PRL103p256404,Starkshift,Nature477p574,Batalov2009}, i.e. an optical $\pi-$pulse. This is the final $\hat{X}$ portion of the map Eq. (\ref{map}). 

\section{Detailed model of experimental protocol}
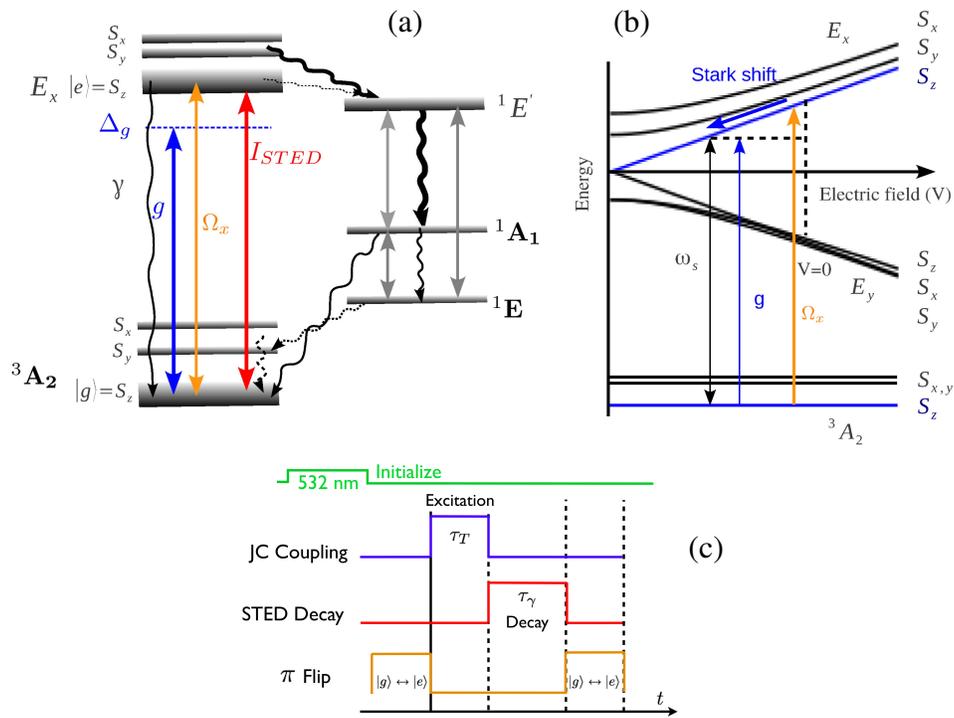
\begin{figure}[htbp]
\begin{center}
\setlength{\unitlength}{1cm}
\begin{picture}(8.5,12)
\put(-2,4){\includegraphics[width=7cm]{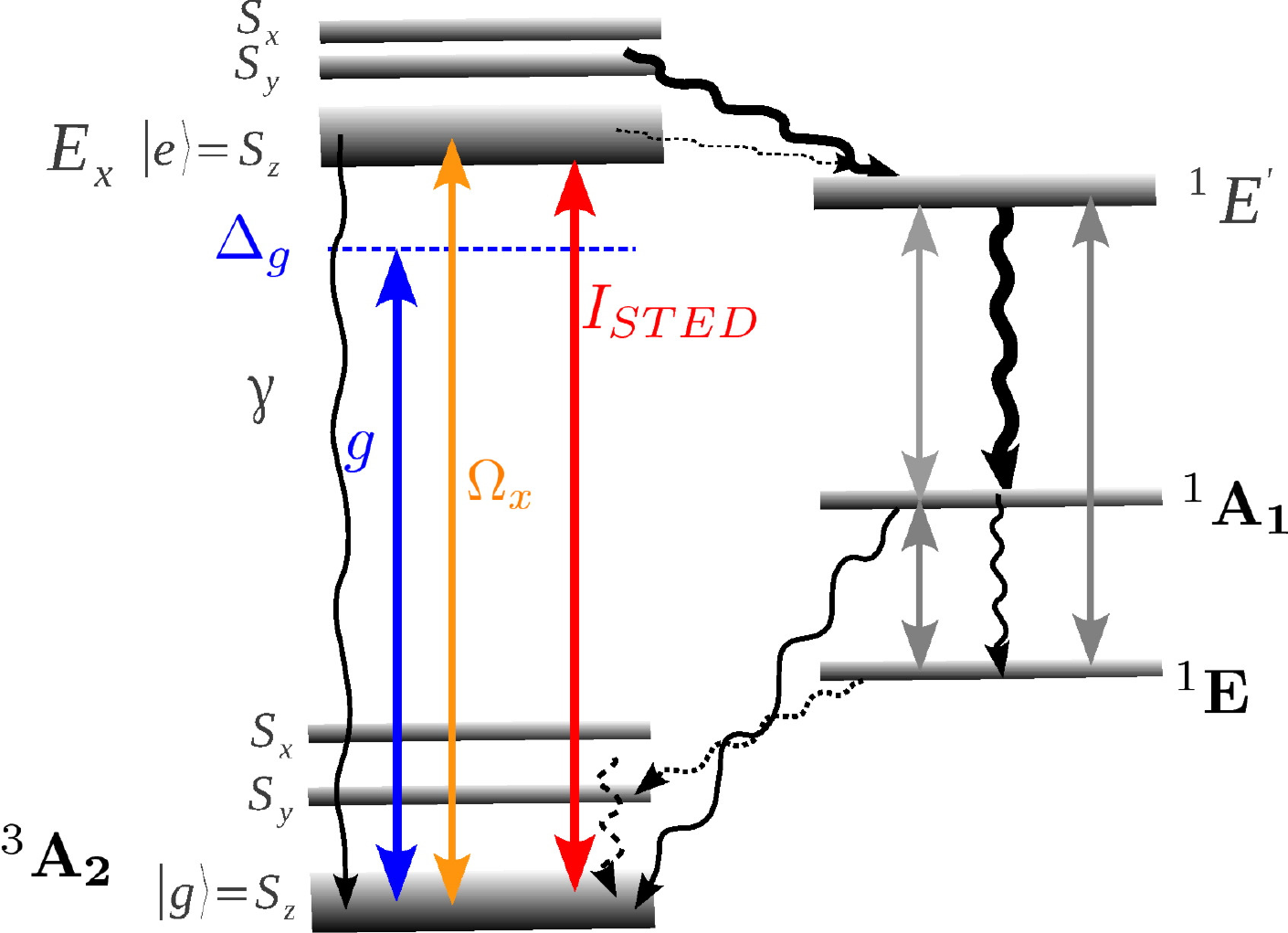}}
\put(5.5,3.5){\includegraphics[width=5cm]{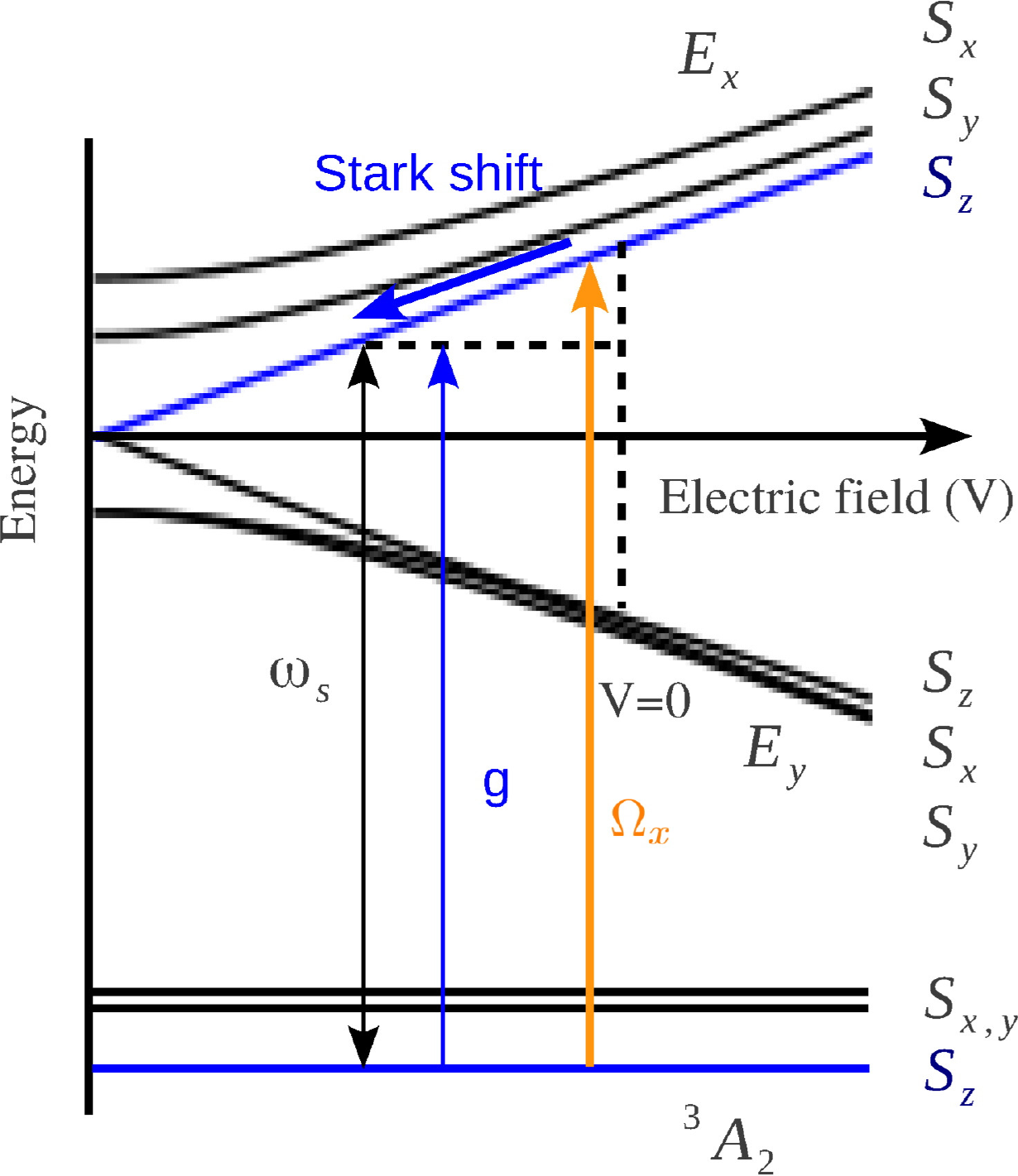}}
\put(1,-1){\includegraphics[width=6.5cm]{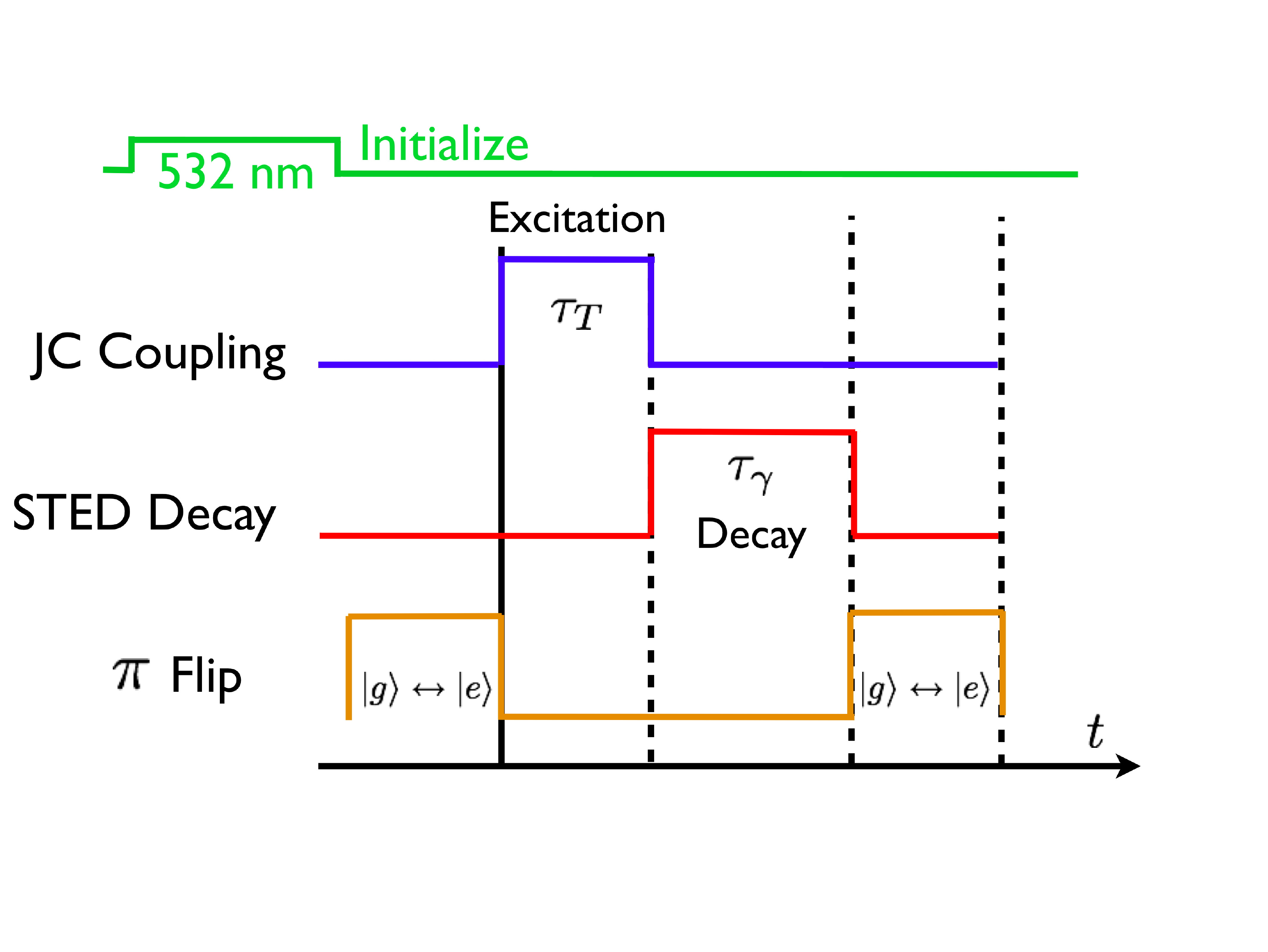}}
\put(3,9){\large (a)}
\put(6,9){\large (b)}
\put(7,2){\large (c)}
\end{picture}
\end{center}
\caption{(a) Level diagram of a NV center showing spin-triplet ground and excited states, as well as the singlet system involved in intersystem crossing \cite{PRB82p201202R,PRB81p041204,NJP13p025025}. Another triplet excited state ${\bf E}_y$ is not shown here.  Also shown is the JC coupling (blue) $g$, decay rate from $|e\rangle$ to $|g\rangle$, the STED illumination (red), and the laser transition for the $\pi$ flip generated by $\Omega_x$ (orange). (b) Eigenvalues of the excited state triplet as a function of applied SEF \cite{NJP10p045004}. The vertical dashed line at $V=0$ marks the splitting due to the strain. At this position the NV$^-$ center can be excited resonantly by a $\pi$ laser pulse $\Omega_x$. An electric field is applied to bring the NV center into resonance with the symmetric mode (resonant frequency $\omega_s$). (c) Time sequence for generation of photonic Fock state showing the initialisation, JC coupling, Decay and $X$ flip.}\label{fig:NVLevel}
\end{figure}

We propose to implement the JC Damped QRW Fock state synthesis using a NV cavity-QED setup. The relevant energy level scheme for the NV center is shown in Fig.~\ref{fig:NVLevel}(a). 
We neglect both the hyperfine electron-nuclear spin coupling and the weak electronic spin-spin interaction \cite{Nature478p221}. The center has an optically allowed transition between an orbital ground state ${^3{\bf A}_2}$ and an orbital excited state ${}^3{\bf E}$. Both the ground and excited states are $S=1$ spin triplets. The ground state has $^3A_2$ symmetry and is split into an $S_x,S_y$ doublet $2.87$${\rm GHz}$~above an $S_z$ singlet due to the zero-field splitting \cite{NJP10p045004}.
The lifetime of excited state ${}^3{\bf E}$ is about $11.6$ ${\rm ns}$, corresponding to a decay rate $\gamma=14$ ${\rm MHz}$ {\cite{PRL97p083002}. Low temperature transform-limited single photon emission spectra from individual NV defects \cite{PRL97p083002} indicates that dephasing is negligible at these low temperatures. Further low temperature experiments demonstrate that the optical excited states of the NV can be isolated from the effects of the nearby phonon sidebands \cite{Batalov2009} at low temperatures.
Throughout our operation below the non-radiative decay to the intersystem state from $|{\bf E}_x,S_z \rangle$ is taken to be negligible. Such intersystem crossing decay contributes to an effective decay to the singlet ground state $S_z$.}

At low temperature (T$<10$ ${\rm K}$), strain in the nanodiamond causes the excited state $^3\bf{E}$ to split into an orbital upper branch ${\bf E}_x$ and an orbital lower branch ${\bf E}_y$ (see Fig.~\ref{fig:NVLevel}(b)). Each branch is a spin triplet formed by three spin states $S_x,S_y$ and $S_z$. The sublevel $|{\bf E}_x,S_z \rangle$ is well isolated from the other five sublevels by several ${\rm GHz}$. Actually the state $|{\bf E}_y,S_z \rangle$ can be isolated from $|{\bf E}_x,S_z \rangle$  because these two sublevels are associated to orthogonal transition dipoles \cite{PRL103p256404,Batalov2009}. Therefore the spin-conserving transition $|^3{\bf A}_2,S_z\rangle \leftrightarrow |{\bf E}_x,S_z \rangle$,  
can be excited resonantly at low strain \cite{Nature477p574,PRL103p256404,Batalov2009}. To suppress further any small spin mixing and phonon-induced transitions within these two excited states \cite{PRL103p256404,Batalov2009}, our setup works with low-strain NV$^-$ centers at cryogenic temperatures. Moreover, the JC coupling $g$ is assumed to be much larger than the decay rate $\gamma$ and the thermal orbital coupling and relaxation rates. In our protocol the duration when the NV center is excited into state $|e\rangle$ state is small and thus the spin mixing can be neglected \cite{PRL103p256404,Batalov2009}. 
Our protocol  primarily involves the transition between $|g\rangle \leftrightarrow |e\rangle$, ($|^3{\bf A}_2,S_z\rangle \leftrightarrow |{\bf E}_x,S_z \rangle$) and excludes other sublevels \cite{Nature477p574}. Thus we are able to treat the NV as a two level system with a transition in the optical  $\sim$ 637${\rm nm}$.

 
The dynamics of our system is given by the Hamiltonian $\hat{H}$ in the rotating wave approximation (RWA) 
\begin{equation}
\begin{split}
  \hat{H} = & \hat{H}_s +\hat{H}_a +\hat{H}_x \,,\\
  \hat{H}_s =& -\hbar (\Delta_g + \Delta_s(t)) \hat{A}_s^\dag \hat{A}_s + \hbar  g [\hat{A}_s^\dag \sigma_- + H.c.] \,,\\
  \hat{H}_x = & \hbar \Omega_x(t)[\hat{\sigma}_- + H.c.]\,,
\end{split}
\end{equation}
where $\hat{\sigma}_-= |g\rangle \langle e|$. $\Delta_{g}=\omega_{zpl}- \omega_s$  is the detuning of the mode $\hat{A}_{s}$ and the ZPL  transition between states $|g\rangle$ and $|e\rangle$ (with frequency $\omega_{zpl}$), in the absence of any Stark shift. $\omega_s=\omega_c+h$ is the resonant frequency of mode $\hat{A}_{s}$ shifted by the scattering $h$. $g$ is the JC coupling strength between a single NV center and a single photon in the cavity. In our scheme, the NV center only couples to the symmetric mode $\hat{A}_s$. This is reasonable because this coupling can be predominately to mode $\hat{A}_s$ by specialized positioning of the nanodiamond \cite{Science319p1062}, and the scattering can also introduce a large detuning between the unwanted mode $A_a$ and the ZPL of NV center.  The Stark shift $\Delta_s(t)$ is used to dynamically control the creation of cavity photons by the NV center due to the JC coupling \cite{Starkshift}. For $\Delta_s(t)=0$, the cavity decouples from the NV center because of the large detuning. This process can be considered as ``the JC coupling off''. The excitation of cavity is turned on if ``the JC coupling on'', i.e. $ \Delta_s(t)=- \Delta_g$. According to our numerical simulation, a fast relaxation from $|e \rangle$ to the ground state $|g\rangle$ is required following the JC coupling phase for the preparation of a high-number Fock state with a high fidelity.  
Here we make use of the concept of ``Stimulated Emission Depletion''  (STED) to dynamically enhance the relaxation process \cite{NaturePhoton3p144} during the ``decay phase'' of the map (see Eq. (\ref{map}) and Fig. \ref{fig:NVLevel}(c)). When the STED beam is applied, the stimulated emission rate $\gamma_{STED}$ becomes to $I_{STED} \gamma/I_s$, with $I_{STED}/I_s$ denoting the ratio of the  STED pulse intensity $I_{STED}$ and the saturation intensity $I_s$. For a lifetime $11.6$~${\rm ns}$, $I_s$ is $\sim 1.85$~${\rm MW}$ ${\rm cm}$$^{-2}$ \cite{NaturePhoton3p144} if a continuous wave (cw) STED beam is applied. A cw STED beam of $20$~${\rm GW}$ ${\rm cm}$$^{-2}$ can enhance the decay rate by four orders of magnitude. During the initial ``JC coupling on'' phase (see Fig. \ref{fig:NVLevel}(c)), we turn off the STED beam and the nominal decay rate of NV center remains $\gamma \sim 14$~${\rm MHz}$.

We now describe the detailed steps in synthesising the process described in Eq. (\ref{map}). We assume that the optical modes in both the cavity and waveguide are initially in the vacuum state yielding initially zero photon number in the cavity. The time sequence for creating a photonic Fock state is shown in Fig.~\ref{fig:NVLevel}(c): Initially the NV center is in its ground state $|g\rangle$ after a short $532$ ${\rm nm}$~ laser pulse optically prepares the defect into the $m_s=0\, (|{}^3A_2,S_z\rangle)$ state. A $\pi$ laser pulse $\Omega_x$ is used to resonantly pump it to the excited state $|e\rangle$. Then the ZPL is tuned  on-resonance with the mode $\hat{A}_s$ to enable the JC coupling $g$ using a SEF. After time $\tau_T$, we turn off the JC coupling but use the STED laser beam to create a fast decay channel to the ground state $|g\rangle$. 
Waiting for time $\tau_{\gamma}$, almost all population decays to the ground state $|g\rangle$ from $|e\rangle$. Then a further  optical $\pi$ pulse generated by $\Omega_x$ is applied to resonantly excite the NV center to $|e\rangle$ again. We repeat these operations until the target state is trapped.

When $\Delta_g + \Delta_s=0$, the NV center resonantly couples to the cavity mode $\hat{A}_s$. The dynamics of the system can then be described by a unitary time evolution operator  and we further now assume that the time duration of this unitary may not be precisely controlled, i.e. we assume some noise in the target JC coupling time $\tau_T$. More precisely we take  $U_{JC}=e^{-i\tau_T (1+\delta \tau) H_s/\hbar}$, where $\delta \tau$ is a normally distributed additional noise in timing with a standard deviation given by a parameter $\sigma_n$. The excitation probability of the cavity to state $|n+1\rangle$ when the NV is in the excited state is thus now given by $P_g=\sin^2 [g \tau_T\,\sqrt{n+1}\,(1+\delta\tau)]$.
Once the SEF is turned off, the STED laser pulse is switched on. The NV center is decoupled from the cavity and relaxes to its GS $S_z$. The timing noise during the decay is not considered in $\tau_\gamma$ because this damping process is insensitive to the timing error. The population in the excited state $|3\rangle$ decays at an effective rate $\gamma_{STED}$ to the ground state $|g\rangle$. Such process can be described by a supperoperator $\varepsilon$ as \cite{QUTool}
\begin{equation}\label{eq:decay}
{\mathcal{E}}\,[ \hat{\rho}]=e^{\left(\hat{\hat L}-i \hat{\hat H}_s/\hbar \right) \tau_\gamma}\,\hat{\rho} \,,
\end{equation}
where the superoperators are defined as $\hat{\hat H}_s \hat{\rho}=[\hat{H}_s,\hat{\rho}]$, 
$\hat{\hat L}\hat{\rho}=\gamma_c/2 (2 \hat{a}\hat{\rho} \hat{a}^\dag -\hat{a}^\dag \hat{a} \hat{\rho} -\hat{\rho} \hat{a}^\dag \hat{a}) +\gamma_{STED}/2 (2\hat{\sigma}_- \hat{\rho} \hat{\sigma}_+ -\hat{\sigma}_z \hat{\rho} -\hat{\rho} \hat{\sigma}_z)$ 
with $\hat{\sigma}_z=|e\rangle \langle e|-|g\rangle \langle g|$ and the density matrix $\hat{\rho}$.
After a time $\tau_\gamma=5\gamma_{STED}^{-1}$, the GS $S_z$ is polarized more than $99\%$ again. The flip $\pi$ laser pulse generated by $\Omega_x$ ($\lambda_x\approx 637$ ${\rm nm}$) turns on successively to flip the NV center to the ES $S_z$. 
We define a flip operator $X=e^{-i\pi(1+\delta x) \sigma_x /2}$ with $\sigma_x=\sigma_+ + \sigma_-$ to model this flip process as $\hat{X}\hat{\rho} \hat{X}^\dag$. $\delta_x$ is a noise having the same statistic property but independent of $\delta_\tau$. Then the density matrix after $l+1$ steps is determined by a recurrence relation 
\begin{equation} \label{eq:PhotonFock1}
\hat{ \rho}_{l+1} = \hat{ X} {\mathcal E}_{SE}\, [\hat{U}_{JC}\, \hat{ \rho}_l\, \hat{ U}_{JC}^\dag]\, \hat{ X}^\dag \,.
\end{equation}
The system is initialized in the state 
$\rho_0=
|e\rangle\langle e|\otimes \rho^c_{vac}$, where $\rho^c_{vac}$ is the density matrix of vacuum state of cavity mode.

\section{Results}
Next we discuss the generation of the photonic Fock state. Throughout our simulation below, we neglect the excitation of $\hat{A}_a$. This requisite can be satisfied by positioning the NV center at the antinode of $\hat{A}_s$ \cite{Science319p1062} or introducing a large scattering. 

To control the JC coupling, we consider a setup shown in Fig.~\ref{fig:setup}, in which the cavity is designed to be off-resonance with the transition $|g\rangle \leftrightarrow |e\rangle$ such that $|\Delta_g| \gg |g|$. A detuning of $\Delta_g=10g$ is large enough to decouple the cavity from the NV center. To switch on the JC coupling, the transition $|g\rangle \leftrightarrow |e\rangle$ is tuned to be on resonance with the cavity, i.e.  $\Delta_g +\Delta_s=0$, by the Stark shift $\Delta_s$ induced by the SEF \cite{Starkshift}. 
However during the JC step the NV center will decay to its ground state $|g\rangle$ at a rate given by $\gamma$ and this decay during the JC step decreases the fidelity of the target Fock state. We assume a static, larger JC coupling $g=30\gamma$ to suppress this detrimental process. By assuming a good optical cavity with $\gamma_c \ll \gamma$, in combination with the large JC coupling strength further improves the ultimate fidelity of the trapped photon state.
\begin{figure}
\begin{center}
\setlength{\unitlength}{1cm}
\begin{picture}(8.5,5)
\put(-2.2,-.4){\includegraphics[width=5.8cm]{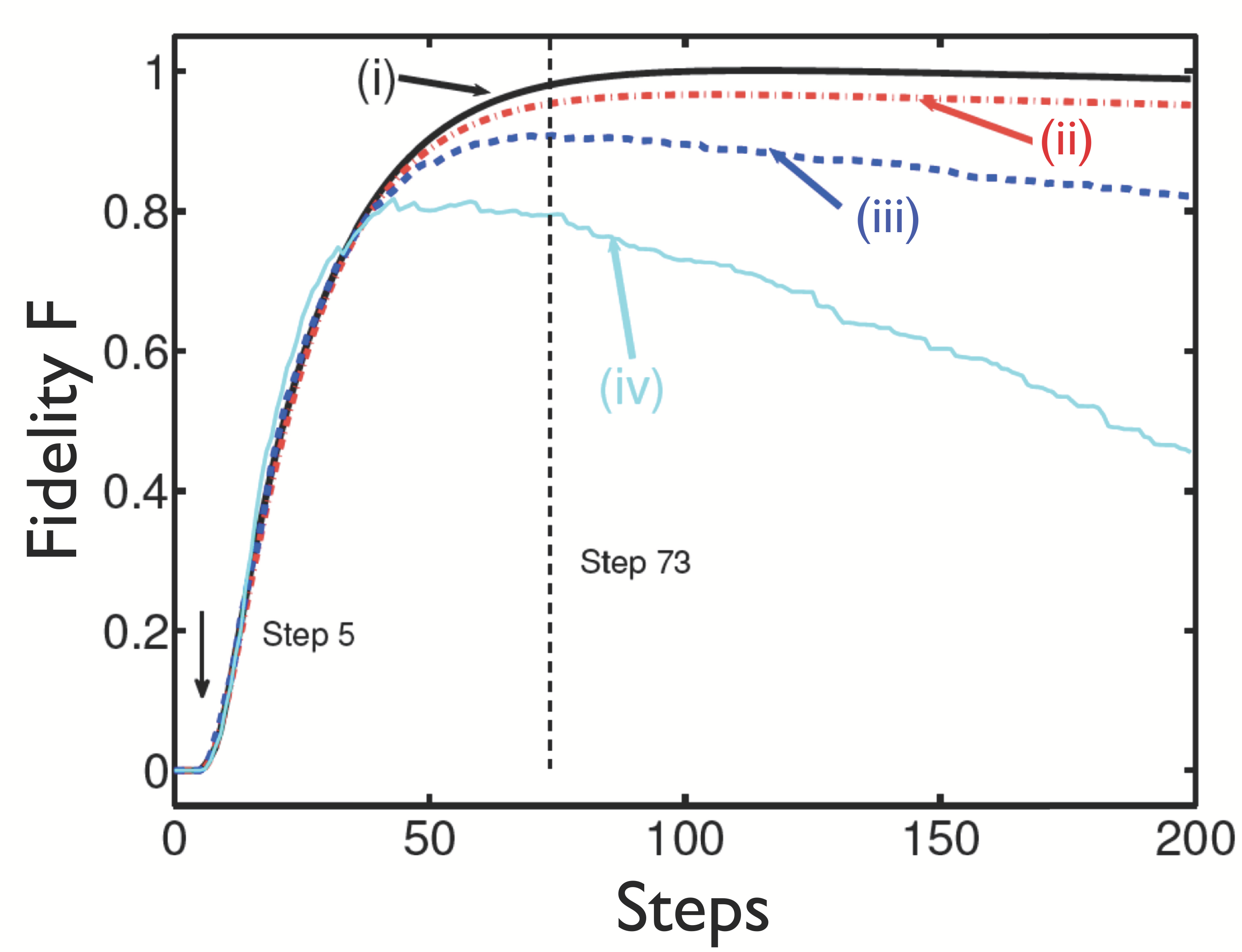}}
\put(3.7,-.2){\includegraphics[width=7.0cm]{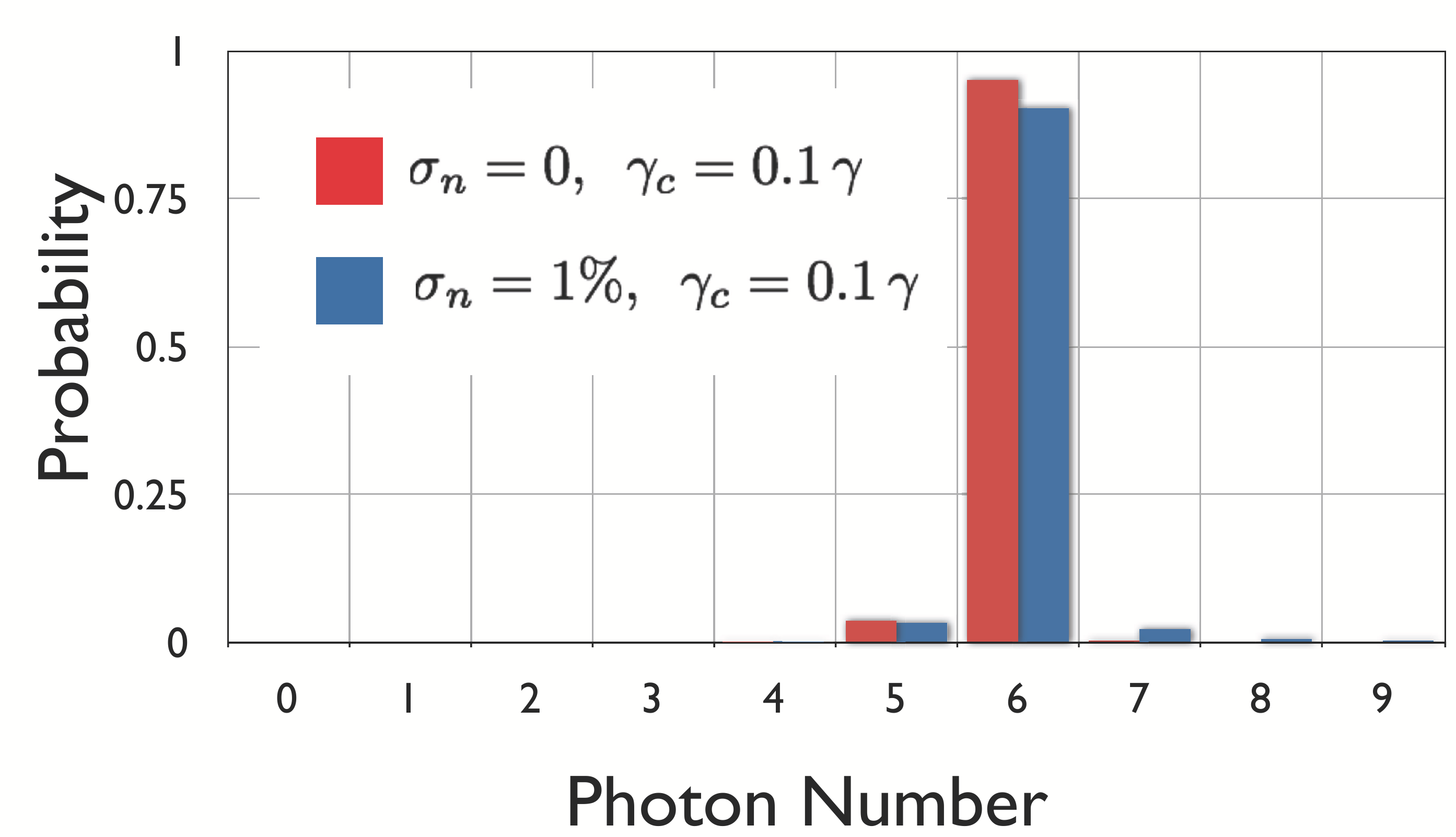}}
\put(2.3,-.28){(a)}
\put(9.4,-.18){(b)}
\end{picture}
\end{center}
\caption{(a) Time evolution of fidelities of target Fock state $|n=6\rangle$.  (i) $\sigma_n=0,\gamma_c=0$; (ii) $\sigma_n=0,\gamma_c=0.1\gamma$; (iii) $\sigma_n=1\%,\gamma_c=0.1\gamma$; (iv) $\sigma_n=2\%, \gamma_c=0$. (b) Probabilities of photon number states at step $73$. Red bar for only cavity decay $\sigma_n=0, \gamma_c=0.1\gamma$; blue bar for $\sigma_n=1\%, \gamma_c=0.1\gamma$. Other parameters are $g=30\gamma,\Delta_g=300\gamma, \gamma_{STED}=10^4\gamma$.}\label{fig:result}
\end{figure}

In the ideal case of no cavity decay, a complete switching on/off JC coupling and no timing error, one can stably trap a Fock state $|n_T\rangle$ with unit fidelity, see the black solid line (i) in Fig.~\ref{fig:result}(a) for instance.

Now we discuss the influences of timing error and the decay of cavity on the generated trapped photonic Fock state taking as an example $n_T=6$.
Before the fifth step, only the Fock states with $|n<6\rangle$ are excited, because in each step the prepared photon number state can only excite the next one. As the operation continues, the target state $|n_T=6\rangle$ is essentially populated. It can be clearly seen from Fig.~\ref{fig:result}(a) that the fidelity $F={\text Tr}[\hat{\rho}\hat{\rho}_{T}]$  with $|n_T=6\rangle$ \cite{PRL93p130501}  increases quickly at first because the excitation probability $P(g,\tau_T,n_T=6)$ is large when the population in $|n=6\rangle$ is small. 
As more population transfers to $|n=6\rangle$, the probability to excite this state decreases. Overall we have found the generation of Focks states is fairly robust with the number of repetitions of the process Eq. (\ref{map}),  providing the timing error is not too large (ii and iii). The cavity mode becomes stable after $\sim 73$ steps.

If only the cavity decay is included (red dashed-dotted line (ii) in Fig.~\ref{fig:result}(a)), the target state is stable once it is prepared after about $100$ steps, and then the fidelity $F$ is very high, about $0.97$. The loss of cavity photon cancel the small probability of pumping from $|n=5\rangle$ to $|n=6\rangle$ if $F$ is large, and consequently leads to the reduce of fidelity. Thus the excitation of state $|n=5\rangle$ is considerable, see red bar in Fig.~\ref{fig:result}(b). 

We notice that the timing error in the JC coupling causes a leakage of the population to higher photon number states. This leakage results in a reduction of the fidelity as the operation continues (iii and iv). 
To provide a limit for the fidelity of a Fock state we can prepare with $n_T \leq 6$, we perform the simulation including both kinds of imperfection: the timing error ($\sigma_n=1\%$) and the cavity decay $\gamma_c=0.1\gamma$. In this case, the probability of $|n=6\rangle$ is about $0.9$ from step $64$ to $94$ (blue dashed line (iii) in Fig.~\ref{fig:result}(a)). Obviously, the prepared Fock state is stable within a wide operation step range. This is an advantage of the trapping state \cite{PRL86p3534}. However about $6\%$ population leaks to the higher photon number state (blue bar in Fig.~\ref{fig:result}(b)). The fidelity gradually reduces as the Q factor of cavity decreases. Hailin Wang's group has demonstrated a microspherical cavity with $\gamma_c \sim 0.4\gamma$ coupling to a nanodiamond \cite{NanoLett6p2075}. Using this number, our simulation shows that the fidelity can still be $0.88$ if $\sigma_n=1\%$.

A timing error of $2\%$ substantially destroys the trapping condition (iv) and causes a considerable excitation of higher state. As a result, a considerable part, about $15\%$, of population leaks to higher photon number states. The fidelity of the target state $|n=6\rangle$ decreases fast from a maximum $F=0.81$ after $60$ steps. However if the operation stops at step $60$, one still can obtain the Fock state $|n=6\rangle$ with high probability. A large error in the interaction time is the crucial reason why a train of atoms successively entering a cavity can not trap a Fock state with high probability \cite{NJP6p97,PRA36p744}. Thanks to a solid-state setup, this timing error can be much smaller during our operation. In contrast, our setup can generate a higher Fock state with a higher fidelity.

\begin{figure}
 \begin{center}
  \includegraphics[width=7cm]{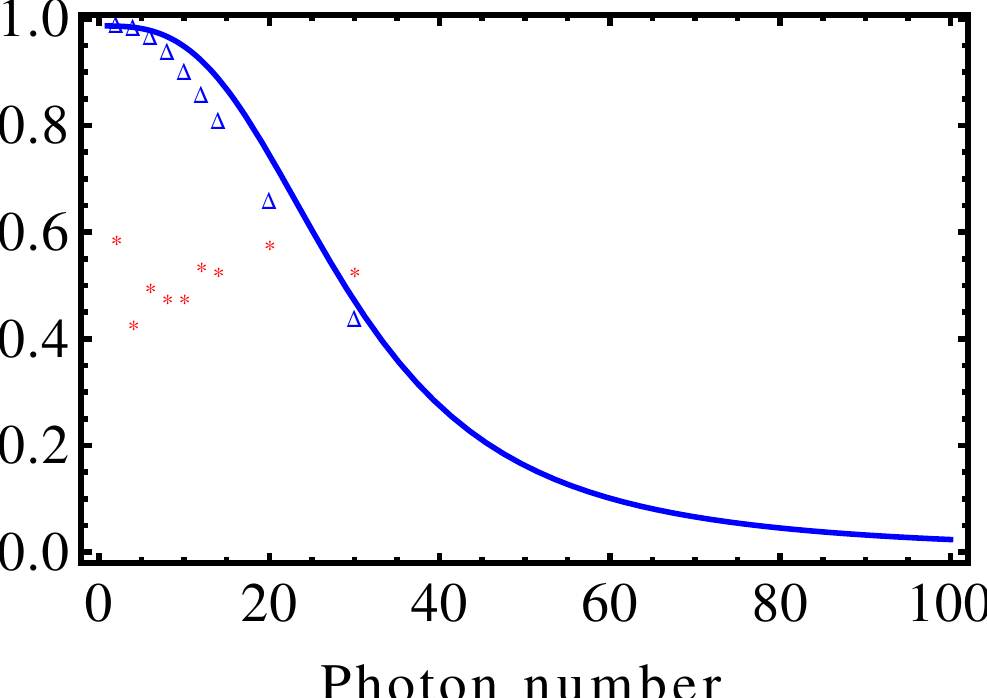}
 \end{center}
\caption{Numerical proof of Relation Eq.~(\ref{eq:FN}). Blue solid line shows the available fidelity $F$ evaluated by Eq.~(\ref{eq:FN}) as a function of target state $|n_T\rangle$ for $\alpha=0.5$ and $M=5, \gamma_q/\gamma_c=10^5$. Blue triangle marks the numerical results for $n_T=2,4,6,8,10,12,14,20,30$ and $\sigma_n=0,\gamma_c=0.1\gamma$. Here $\gamma_q$ is equal to $\gamma_{STED}$. Numerical evaluation of $\alpha$ is marked as red stars. Other parameters are $g=30\gamma, \Delta_g=300\gamma$.}\label{fig:FN}
\end{figure}
Even if the timing of JC interaction can be controlled perfectly, the decay of the cavity also limit the available number of photon of Fock state for a set fidelity $F$. Equation~(\ref{eq:FN}) provides a good estimation for the maximum of $n_T$ if $F$ is set. The constant $\alpha$ (about $0.5$) is numerically evaluated for $\sigma_n=0,\gamma_{STED}/\gamma_c=10^5$. Using this value in Eq.~(\ref{eq:FN}), the fidelity for a certain target state $|n_T\rangle$ is shown in Fig.~\ref{fig:FN}. Clearly, the estimation given by Eq.~(\ref{eq:FN}) is consistent with the numerical results.

To generate a Fock state $|n=6\rangle$, we need $\gamma_{STED} > 10^4 \gamma_c$. In the presence of noise, the decay rate $\gamma_{STED}$ need be larger.
To perform these simulations we used a cavity with $Q\sim 3\times 10^8$ \cite{NaturePhoton6p369,Nature424p839,OL23p247,OL21p453,PRA74p063806,OE15p3390}, corresponding to a decay rate of $\gamma_c\sim 2\pi \times 1.4$~${\rm MHz}$.
The nanodiamond embedded in the cavity contributes an extra loss channel to the cavity and subsequently reduce the Q factor. However this induced loss is proportional to $r^6$, where $r$ is the radius of particle \cite{PRL99p173603,NPhoton4p46}. This contribution of loss is negligible if $r<10$~nm and nanodiamonds containing nitrogen vacancy centres in such small nano diamonds have been made \cite{NPhoton7p11,Small5p1649}. Experiments have demonstrated that the Q factor of a cavity embedding a nanoparticle, such as a nanodimaond \cite{PRA74p063806}, or potassium chloride nanoparticle \cite{NPhoton4p46}, can be larger than $3\times 10^8$. The nanodiamond also causes scattering in the cavity and leads to a doubling of the linewidth of the cavity or mode splitting. This scattering rate decreases quickly ($\propto r^3$) as the size of particle decreases \cite{PRL99p173603,NPhoton4p46}. On the other hand we use the nanodiamond to selectively excite the symmetric mode. Therefore the effect of scattering on the generation of the target state can be neglected for $r<10$~nm. 
We use the typical value $\gamma /2\pi=14$~${\rm MHz}$~ for the decay rate of the excited state $|3\rangle$ of a single NV center \cite{PRL97p083002,PRL105p177403}. This decay rate can be enhanced by four orders in magnitude if a $20$ ${\rm GW}$ ${\rm cm}$$^{-2}$ cw STED beam is applied. To suppress the decay of population from the state $|3\rangle$ during the JC coupling on, we need a JC coupling strength $g=30\gamma \sim 400$  ${\rm MHz}$, which can be reached in the current experiments \cite{OE17p8081}. This large value of $g$ allows for shorter $\tau_T$ and on these times scales the  mixing between states ${\bf E}_x$ and ${\bf E}_y$ is negligible. For this coupling strength, a Stark shift of $\Delta_s=10g \sim 2\pi \times 4$ ${\rm GHz}$~ is large enough to switch on/off the excitation of cavity. Such Stark shift can be created using two electrodes separated by $10$ ${\mu}{\rm m}$~ and positioned $10$ ${\mu}{\rm m}$~ above the NV center \cite{PRL107p266403}.

\section{Conclusion}
In conclusion, we have proposed a solid-state setup consisting of a single NV$^-$ center and a high-$Q$ toroidal cavity for the generation of a multi-photon optical Fock state through the iteration of a damped Jaynes Cummings quantum random walk. By iterating this walk step we found a method to trap an on-demand photonic Fock state with high fidelity within the cavity.

\section*{Acknowledgments}
One of us (D.E.) is grateful to the Macquarie University Research Centre for Quantum Science and Technology, for hospitality during a sabbatical stay during which this work was initiated. We also acknowledge support from the ARC Centre of Excellence in Engineered Quantum Systems and EU Project Quantip.

\end{document}